\documentclass[twocolumn,english]{revtex4-1}
\usepackage[T1]{fontenc}
\usepackage[latin9]{inputenc}
\usepackage{verbatim}
\usepackage{float}
\usepackage{amstext}
\usepackage{amsmath}
\usepackage{amssymb}
\usepackage{graphicx}
\usepackage{esint}
\usepackage[colorlinks,linkcolor=blue]{hyperref}

\makeatletter
\usepackage{graphicx}
\usepackage{dcolumn}
\usepackage{bm}
\usepackage{hyperref}

\makeatother

\usepackage{babel}
\begin{document}
\preprint{APS/123-QED}

\title{Two-dimensional regular string black hole \\ via complete $\alpha^{\prime}$ corrections}

\author{Shuxuan Ying}
\email{ysxuan@cqu.edu.cn}
\affiliation{Department of Physics, Chongqing University, Chongqing, 401331, China}


\begin{abstract}

In string theory, an important challenge is to show if the singularity of black holes can be smoothed out by the complete $\alpha^{\prime}$ corrections. The simplest case is to consider a 2D string black hole or 3D black string. This problem was discussed in a gauged Wess-Zumino-Witten (WZW) model and the results are supposed to be correct to all orders in $\alpha^{\prime}$ corrections. Based on the recent remarkable progress on classifying all the $\alpha^{\prime}$ corrections, in this work, we re-study this problem  with the low energy effective spacetime action, and  provide  classes of exact non-perturbative and non-singular solutions of the 2D black hole/3D black string via complete $\alpha^{\prime}$ corrections.

\end{abstract}

\maketitle

How to resolve a singularity of a black hole is a long-standing problem
in general relativity. It is well-known that the black hole possesses
two kinds of infinities: one is the coordinate singularity which can be removed by
coordinates transformations;
another is the curvature singularity which cannot be smoothed out and is unavoidable in Einstein's
gravity \cite{Penrose:1964wq,Hawking:1965mf}. The regular black hole
solutions must satisfy two requirements: I) The curvature invariants
(such as Kretschmann scalar) are regular everywhere; II) The geodesics
are complete. However,  there is yet
no systematic methods to obtain  regular solutions from solving
Einstein's equation exactly without any ad hoc settings. People
therefore expect a theory of quantum gravity will provide inspirations
to solve this problem completely. String theory as one candidate of
quantum gravitational theories should be able to answer this question.

The simplest black hole solution in string theory, namely 1+1 dimensional
string black hole, was obtained by solving a 2D closed string's low
energy effective action \cite{Mandal:1991tz,Lemos:1993py}. This solution is only
valid when the string length scale $\sqrt{\alpha^{\prime}}$ is small
compared to the radius of spacetime curvature. One may wonder how
to obtain the exact string black hole solutions of the full action,
and whether the spacetime singularity disappears in these solutions.
However, it looks impossible since the low energy effective action
with complete $\alpha^{\prime}$ corrections is unknown. As an alternative,
Witten obtained the exact metric which described the region outside
the event horizon of 2D string black hole through the $SL\left(2,R\right)/U\left(1\right)$
gauged WZW model, this result is conformally invariant to all orders
in $1/k$ when $k\rightarrow\infty$ ($k\sim1/\alpha^{\prime}$ is
the Kac--Moody level) \cite{Witten:1991}. This metric still possesses
a curvature singularity in the maximally extended
spacetime. Then, in ref. \cite{Dijkgraaf:1991ba}, Dijkgraaf, Verlinde
and Verlinde discovered the exact 2D string black hole for general
$k$, which was supposed to be correct to all orders in $\alpha^{\prime}$.
Based on this work, Perry and Teo \cite{Perry:1993ry}, and Yi \cite{Yi:1993gh}
studied its maximally extended spacetime.  The Weyl invariance
of this solution is then verified by Tseytlin up to 3-loops ($\alpha^{\prime2}$)
in the bosonic sigma model\cite{Tseytlin:1991ht}.
The supersymmetric 4-loops ($\alpha^{\prime3}$) was also checked in
ref. \cite{Jack:1992mk}.

In recent works \cite{Hohm:2015doa,Hohm:2019ccp,Hohm:2019jgu}, Hohm
and Zwiebach reconsidered how to classify all orders $\alpha^{\prime}$
corrections of the low energy effective action. This progress makes
it possible to re-study the exact 2D string black hole systematically.
Hohm and Zwiebach's motivation was based on two reasons: I) the tree-level
string effective action and its first order $\alpha^{\prime}$ correction
can be put into an explicit $O\left(d,d\right)$ covariant form by
suitable field redefinitions \cite{Veneziano:1991ek, Sen:1991zi,Sen:1991cn, Meissner:1991zj,Meissner:1996sa};
II) Sen proved that for configurations independent of $m$ coordinates,
all orders in $\alpha^{\prime}$ expansion possess the $O\left(m,m\right)$
symmetry \cite{Sen:1991zi,Sen:1991cn}. Hohm and Zwiebach therefore
assumed that the standard $O\left(d,d\right)$ matrix always keep
its form unchanged to all orders in $\alpha^{\prime}$, meanwhile
the $O\left(d,d\right)$-breaking terms could be absorbed into the
standard matrix by the field redefinitions. With this assumption,
Hohm and Zwiebach showed that all orders in $\alpha^{\prime}$ are
classified by even powers of Hubble parameter in FLRW cosmological
background. The dilaton appears trivially and only first order time
derivatives need to be included. Since the equations of motion (EOM)
only include first two derivatives of spacetime metric, the theory
is ghost free and exactly solvable. This significant progress makes
it possible to study non-perturbative stringy effects. Based on this
work, we have found that the big-bang singularity could be smoothed
out by the complete $\alpha^{\prime}$ corrections \cite{Wang:2019kez,Wang:2019dcj}.
It is also proved that the Hohm and Zwiebach's derivation can also
apply to the domain wall background \cite{Wang:2019mwi}, so that
the naked singularities can be resolved \cite{Ying:2021xse}. Based on these attempts,
it is reasonable to believe that the  $\alpha^{\prime}$ corrections will resolve the singularity
of black hole \cite{Bars:1992sr}. However, the process is
tricky for the general spherically symmetric black holes, because they break the $O\left(d,d\right)$
symmetry and Hohm-Zwiebach action cannot be derived from this background.
As an alternative, we consider the 2D black hole/3D black string at first.

In this paper, our aim is to find regular solutions which exactly
solve the EOM of completely $\alpha^{\prime}$ corrected closed string
theory. In the perturbative region
$\alpha^{\prime}\rightarrow0$, it reduces to traditional 2D black
hole/3D black string. At first, let us start by the 2D low energy effective action
of closed string:

\begin{equation}
S=\int d^{2}x\sqrt{-g}e^{-2\phi}\left(R+4\left(\nabla\phi\right)^{2}+\lambda^{2}\right),\label{eq:original action}
\end{equation}

\noindent where $g_{\mu\nu}$ is the string metric, $\phi$ is the
physical dilaton, $\lambda^{2}=-\frac{2\left(D-26\right)}{3\alpha^{\prime}}$
and we set Kalb-Ramond field $b_{\mu\nu}=0$ for simplicity.
This two dimensional gravity has dynamics due to a pre-factor $e^{-2\phi}$.
The black hole solution of this action is given by \cite{Mandal:1991tz}:

\begin{eqnarray}
ds^{2} & = & -\left(1-\frac{M}{r}\right)dt^{2}+\left(1-\frac{M}{r}\right)^{-1}\frac{1}{\lambda^{2}r^{2}}dr^{2},\nonumber \\
\phi & = & -\frac{1}{2}\ln\left(\frac{2}{M}r\right).\label{eq:3D black string metric}
\end{eqnarray}

\noindent where we set an integral constant $\phi_{0}=0$ in $\phi$
for simplicity. If we add a direction $d\varphi^{2}$ in this metric,
it is called a black string solution and firstly discovered by Horne
and Horowitz \cite{Horne:1991gn}. The difference betwween 2D and 3D theories is only $\lambda$ due to its definition. When we consider the extra dimension $\varphi$, the 3D black string solution is a simple product of $d\varphi^2$ and the two dimensional metric, which does not affect the action \cite{Horne:1991gn}. Therefore, although we only study the 2D black
hole in this paper, our result also applies to the 3D black string. Back to the metric (\ref{eq:3D black string metric})
, the event horizon is located at $r=M$ and its curvature singularity
is $r=0$ due to a scalar curvature $R=\frac{\lambda^{2}M}{r}$.  Keep in mind that there are two kinds of
coordinate transformations which cover different regions of maximally
extended spacetime. The first one is

\begin{equation}
\frac{r}{M}=\cosh^{2}\left(\frac{\lambda}{2}x\right),\label{eq:coordinates transformation}
\end{equation}

\noindent where $r\geq M$ and $x\geq0$. Utilizing this coordinate
transformation, the metric (\ref{eq:3D black string metric}) becomes

\begin{eqnarray}
ds^{2} & = & -\tanh^{2}\left(\frac{\lambda}{2}x\right)dt^{2}+dx^{2},\nonumber \\
\Phi & = & -\ln\left(\sinh\left(\lambda x\right)\right),\label{eq:O(2,2) metric}
\end{eqnarray}

\noindent where $O(d,d)$ invariant dilaton is defined by

\begin{equation}
\Phi=2\phi-\ln\sqrt{\det g_{ij}}.\label{eq:odd invariant dilaton}
\end{equation}

\noindent This metric is well-known as Witten's 2D black hole solution,
which was obtained by the $SL\left(2,R\right)/U\left(1\right)$ gauged
WZW model \cite{Witten:1991}. This metric could be applied in Hohm-Zwiebach
action directly. However, it does not possess a curvature singularity
since it only describes the region  outside the event horizon
($x=0$), and the scalar curvature $R_{0}=\lambda^{2}\cosh^{-2}\left(\frac{\lambda x}{2}\right)$
is regular in this region. To discover the curvature singularity
of the metric (\ref{eq:3D black string metric}), we need to adopt
the second kind of coordinate transformations:

\begin{equation}
\frac{r}{M}=\cos^{2}\left(\frac{\lambda}{2}x\right),\label{eq:coor t}
\end{equation}

\noindent where $0\leq r\leq M$ and we only consider one period,
namely $0\leq x\leq\frac{\pi}{\lambda}$. Based on this transformation,
the metric (\ref{eq:3D black string metric}) becomes

\begin{eqnarray}
ds^{2} & = & -dx^{2}+\tan^{2}\left(\frac{\lambda}{2}x\right)dt^{2},\nonumber \\
\Phi & = & -\ln\left(\sin\left(\lambda x\right)\right),\label{eq:inner metric}
\end{eqnarray}

\noindent which describes the inner metric of black hole, and $x$
here plays a role as the time-like direction.This metric (\ref{eq:inner metric})
topologically corresponds to an annulus. The event horizon is located at
$x=0$ and the curvature singularity is the boundary
$x=\frac{\pi}{\lambda}$ due to the scalar curvature $R_{0}=\lambda^{2}\cos^{-2}\left(\frac{\lambda x}{2}\right)$.
Our aim is to remove the curvature singularity of (\ref{eq:inner metric})
by the complete $\alpha^{\prime}$ corrections.

Based on (\ref{eq:inner metric}), we first assume the ansatz,

\begin{equation}
ds^{2}=-dx^{2}+a\left(x\right)^{2}dt^{2}.\label{eq:ansatz}
\end{equation}

\noindent The closed string fields which depend on this metric possess
$O\left(1,1\right)$ symmetry. It is worth noting that we can also use the ansatz  (\ref{eq:3D black string metric}), which shares the same result with (\ref{eq:inner metric}) by the coordinate transformation (\ref{eq:coor t}).
Based on this ansatz, Hohm and Zwiebach
showed that the following low energy effective action with complete
$\alpha^{\prime}$ corrections could be rewritten as

\begin{eqnarray}
I_{HZ} & = & \int d^{2}x\sqrt{-g}e^{-2\phi}\left(R+4\left(\partial\phi\right)^{2}\right.\nonumber \\
 &  & \left.+\frac{1}{4}\alpha^{\prime}\left(R^{\mu\nu\rho\sigma}R_{\mu\nu\rho\sigma}+\ldots\right)+\alpha^{\prime2}\left(\ldots\right)+\ldots\right),\nonumber \\
 & = & \int dxe^{-\Phi}\left(-\dot{\Phi}^{2}-\sum_{\text{k=1}}^{\infty}\left(-\alpha^{\prime}\right)^{k-1}2^{2k+1}c_{k}H^{2k}\right),\nonumber \\
\label{eq:Odd action with alpha}
\end{eqnarray}

\noindent where  $\dot{f}\left(x\right)\equiv\partial_{x}f\left(x\right)$,
$H\left(x\right)\equiv\frac{\dot{a}\left(x\right)}{a\left(x\right)}$,
$c_{1}=-\frac{1}{8}$, $c_{2}=\frac{1}{64}$, $c_{3}=-\frac{1}{3.2^{7}}$,
$c_{4}=\frac{1}{2^{15}}-\frac{1}{2^{12}}\zeta\left(3\right)$ and $c_{k>4}$'s
are unknown coefficients for a bosonic case \cite{Codina:2021cxh}.
In addition, the $O(d,d)$ symmetry of the action (\ref{eq:Odd action with alpha})
is manifested by the following transformations:

\begin{equation}
\Phi\rightarrow\Phi,\qquad a\rightarrow a^{-1},\qquad H\rightarrow-H.
\end{equation}

\noindent To match the model (\ref{eq:original action}),
we need to add an $O\left(d,d\right)$ invariant constant
into the action (\ref{eq:Odd action with alpha}):

\begin{equation}
I_{m}=\int d^{2}xe^{-\Phi}\lambda^{2}.
\end{equation}

\noindent which is also a scalar under the general coordinate transformations.
The EOM   then is

\begin{eqnarray}
\ddot{\Phi}+\frac{1}{2}Hf\left(H\right) & = & 0,\nonumber \\
\frac{d}{dx}\left(e^{-\Phi}f\left(H\right)\right) & = & 0,\nonumber \\
\dot{\Phi}^{2}+g\left(H\right)+\lambda^{2} & = & 0,\label{eq:corrected EOM}
\end{eqnarray}

\noindent where

\begin{eqnarray}
f\left(H\right) & = & \sum_{\text{k=1}}^{\infty}\left(-\alpha^{\prime}\right)^{k-1}2^{2\left(k+1\right)}kc_{k}H^{2k-1}\nonumber \\
 & = & -2H-\alpha^{\prime}2H^{3}+\cdots,\nonumber \\
g\left(H\right) & = & \sum_{\text{k=1}}^{\infty}\left(-\alpha^{\prime}\right)^{k-1}2^{2k+1}\left(2k-1\right)c_{k}H^{2k}\nonumber \\
 & = & -H^{2}-\alpha^{\prime}\frac{3}{2}H^{4}+\cdots,\label{eq:EOM fh gh}
\end{eqnarray}

\noindent and there is an extra constraint $\dot{g}\left(H\right)=H\dot{f}\left(H\right)$.
It is easy to check that the solution (\ref{eq:O(2,2) metric}) and
(\ref{eq:inner metric}) satisfy the EOM (\ref{eq:corrected EOM})
at zeroth order in $\alpha^{\prime}$. Furthermore, we wish to stress
that $\alpha^{\prime}$ is any positive real number in the EOM (\ref{eq:corrected EOM})
and the action (\ref{eq:Odd action with alpha}). Therefore, eq. (\ref{eq:EOM fh gh})
are not the simple expansions in the limit $\alpha^{\prime}\rightarrow0$,
but the non-perturbative series in  general $\alpha^{\prime}$.

To remove the curvature singularity of \textcolor{red}{(\ref{eq:inner metric})} to
obtain a non-perturbative and non-singular solution of EOM (\ref{eq:corrected EOM})
with complete $\alpha^{\prime}$ corrections, there are two requirements:
\begin{itemize}
\item The solutions are assumed to be regular everywhere for  general
$\alpha^{\prime}$.
\item In the perturbative regime, namely, as $\alpha^{\prime}\rightarrow0$,
the solutions must reduce to the perturbative solutions of EOM (\ref{eq:corrected EOM}).
Although, we know that $\alpha^{\prime}\lambda^{2}=-\frac{2\left(D-26\right)}{3}=1$6
when $D=2$ which satisfies the condition $\alpha^{\prime}\lambda^{2}>0$,
we treat $\alpha^{\prime}\lambda^{2}$ as a general value here as in
\cite{Dijkgraaf:1991ba}.
\end{itemize}
In the body, we present one class of general non-perturbative solution
which covers all coefficients $c_{k}$. In the Appendix A, we will give
other possible general solutions. Now, let us start by calculating
the perturbative solutions of EOM (\ref{eq:corrected EOM}). For convenience,
we introduce a new variable $\Omega$ as
\begin{equation}
\Omega\equiv e^{-\Phi},\label{eq:pertur notation}
\end{equation}

\noindent where $\dot{\Omega}=-\dot{\Phi}\Omega$ and $\ddot{\Omega}=\left(-\ddot{\Phi}+\dot{\Phi}^{2}\right)\Omega$.
And the EOM (\ref{eq:corrected EOM}) become

\noindent
\begin{eqnarray}
\ddot{\Omega}-\left(h\left(H\right)-\lambda^{2}\right)\Omega & = & 0,\nonumber \\
\frac{d}{dt}\left(\Omega f\left(H\right)\right) & = & 0,\nonumber \\
\dot{\Omega}^{2}+\left(g\left(H\right)+\lambda^{2}\right)\Omega^{2} & = & 0,\label{eq:reEOM}
\end{eqnarray}

\noindent where we define a new function

\begin{equation}
h\left(H\right)\equiv\frac{1}{2}Hf\left(H\right)-g\left(H\right)=\alpha^{\prime}\frac{1}{2}H^{4}+\ldots,
\end{equation}

\noindent It is easy to see that $h\left(H\right)=0$ at the zeroth
order in $\alpha^{\prime}$. Then, we assume the perturbative solutions
of the EOM (\ref{eq:reEOM}) take the following forms when $\alpha^{\prime}\rightarrow0$:

\begin{eqnarray}
\Omega\left(x\right) & = & \Omega_{0}\left(x\right)+\alpha^{\prime}\Omega_{1}\left(x\right)+\alpha^{\prime2}\Omega_{2}\left(x\right)+\ldots,\nonumber \\
H\left(x\right) & = & H_{0}\left(x\right)+\alpha^{\prime}H_{1}\left(x\right)+\alpha^{\prime2}H_{2}\left(x\right)+\ldots,\label{eq:pertur form}
\end{eqnarray}

\noindent where we denote $\Omega_{i}$ and $H_{i}$ as the $i$-th
order of the perturbative solutions. Therefore, the perturbative solution
can be calculated order by order

\begin{eqnarray}
H\left(x\right) & = & \lambda\text{csc}\left(\lambda x\right)-\frac{\lambda^{3}}{4}\frac{\left(\cos\left(2\lambda x\right)+4\right)}{\sin^{3}\left(\lambda x\right)}\alpha^{\prime}+\cdots,\nonumber \\
\Omega\left(x\right) & = & \sin\left(\lambda x\right)+\frac{\lambda^{2}}{4}\frac{\cos\left(2\lambda x\right)}{\sin\left(\lambda x\right)}\alpha^{\prime}+\cdots.\label{eq:perturbed solution}
\end{eqnarray}

\noindent Due to the eq. (\ref{eq:pertur notation}), we also have

\begin{equation}
\Phi\left(x\right)=-\log\left(\sin\left(\lambda x\right)\right)-\frac{1}{4}\lambda^{2}\left(\cot^{2}\left(\lambda x\right)-1\right)\alpha^{\prime}+\cdots.\label{eq:perturbed solution dilaton}
\end{equation}

\noindent Based on the perturbative solutions (\ref{eq:perturbed solution})
and (\ref{eq:perturbed solution dilaton}) up to any higher order,
we can figure out the general non-perturbative and non-singular solution
which exactly solves the EOM (\ref{eq:corrected EOM}) and covers
all coefficients $c_{k}$:

\begin{equation}
\Phi\left(x\right)=\frac{1}{2}\log\left(\frac{\sum_{\text{k=1}}^{N}\left(\alpha^{\prime}\lambda^{2}\right)^{k-1}}{\sum_{\text{k=1}}^{N}\sigma_{k}\left(\lambda x,c_{k}\right)\left(\alpha^{\prime}\lambda^{2}\right)^{k-1}}\right),\label{eq:general solution}
\end{equation}

\noindent where $\sigma_{k}$'s are functions of $\lambda x$ and
$c_{k}$. After obtained regular $\Phi\left(x\right)$, the regularity of
$H\left(x\right)$, $f\left(x\right)$ and $g\left(x\right)$
is guaranteed  due to the EOM (\ref{eq:corrected EOM}).
Moreover, in the perturbative regime $\alpha^{\prime}\rightarrow0$,
the general solution $\Phi\left(x\right)$ is expanded as,

\begin{eqnarray}
\Phi\left(x\right) & = & -\frac{1}{2}\log\left(\sigma_{1}\right)+\frac{\left(\sigma_{1}-\sigma_{2}\right)}{2\sigma_{1}}\alpha^{\prime}\lambda^{2}\nonumber \\
 &  & +\frac{\left(\sigma_{1}^{2}-2\sigma_{3}\sigma_{1}+\sigma_{2}^{2}\right)}{4\sigma_{1}^{2}}\left(\alpha^{\prime}\lambda^{2}\right)^{2}+\cdots.\label{eq:general solution expansion}
\end{eqnarray}

\noindent To analyze the general solution (\ref{eq:general solution}),
we choose the special case whose perturbative expansion only covers
$c_{1}=-\frac{1}{8}$ and $c_{2}=\frac{1}{64}$. It is not difficult
to check that the general solution (\ref{eq:general solution}) which covers the coefficients $c_{N>2}$
does not affect our following argument due to the ref. \cite{Wang:2019dcj}. Therefore, to match the expansion
(\ref{eq:general solution expansion}) with the perturbative solution
(\ref{eq:perturbed solution dilaton}), we can fix the functions $\sigma_{1}=\sin^{2}\left(\lambda x\right)$,
$\sigma_{2}=\frac{1}{2}$
and $N=2$ such that

\begin{eqnarray}
\Phi\left(x\right) & = & \log\sqrt{\frac{1+\alpha^{\prime}\lambda^{2}}{\sin^{2}\left(\lambda x\right)+\frac{1}{2}\alpha^{\prime}\lambda^{2}}},\nonumber \\
H\left(x\right) & = & -\frac{\sqrt{2}\lambda\left(\left(\alpha^{\prime}\lambda^{2}+1\right)\cos\left(2\lambda x\right)-1\right)}{\left(\alpha^{\prime}\lambda^{2}+1\right)^{1/2}\left(\alpha^{\prime}\lambda^{2}+1-\cos\left(2\lambda x\right)\right)^{3/2}},\nonumber \\
f\left(x\right) & = & -2\sqrt{2}\lambda\left(\frac{\alpha^{\prime}\lambda^{2}+1}{\alpha^{\prime}\lambda^{2}+1-\cos\left(2\lambda x\right)}\right)^{1/2},\nonumber \\
g\left(x\right) & = & \frac{\lambda^{2}}{\left(\alpha^{\prime}\lambda^{2}+1-\cos\left(2\lambda x\right)\right)^{2}}\left(-\alpha^{\prime}\lambda^{2}\left(\alpha^{\prime}\lambda^{2}+2\right)\right.\nonumber \\
 &  & \left.+2\left(\alpha^{\prime}\lambda^{2}+1\right)\cos\left(2\lambda x\right)-2\right).\nonumber \\
\label{eq:solution}
\end{eqnarray}

\begin{figure}[H]
\begin{centering}
\includegraphics[scale=0.4]{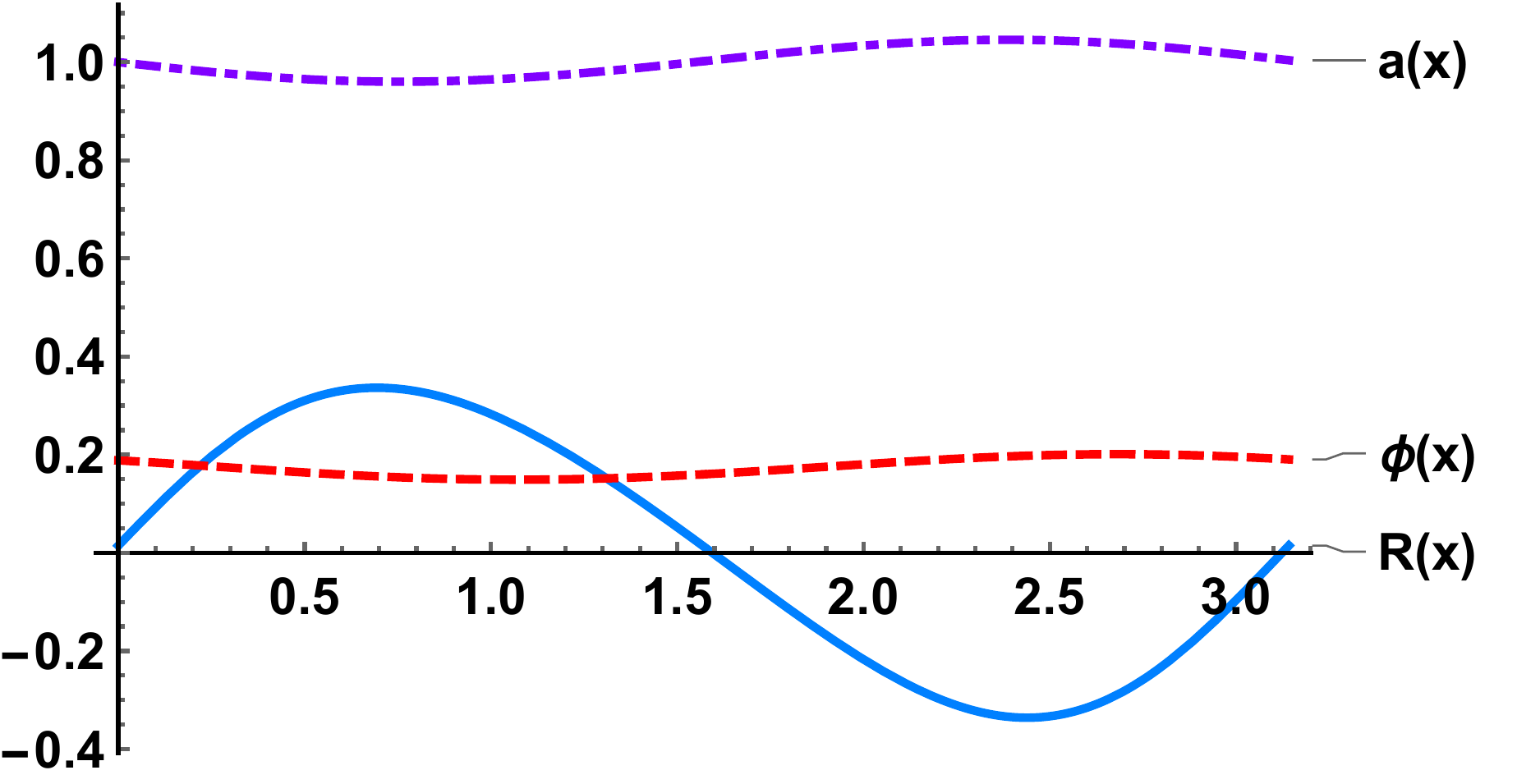}
\par\end{centering}
\centering{}\caption{\label{fig:scalar curvature} The figure of $R\left(x\right)$, $a\left(x\right)$
and $\phi\left(x\right)$ at $\alpha^{\prime}\lambda^{2}=16$.}
\end{figure}

\noindent Based on the ansatz (\ref{eq:ansatz}), Kretschmann scalar
is given by $R_{\mu\nu\rho\sigma}R^{\mu\nu\rho\sigma}=\frac{1}{2}R_{\mu\nu}R^{\mu\nu}=R^{2}=4\left(\dot{H}+H^{2}\right)^{2}$.
Therefore, the regular solution exists when $\alpha^{\prime}\lambda^{2}>0$
in the region $0\leq x\leq\frac{\pi}{\lambda}$. From the solution (\ref{eq:solution}),
we will get

\begin{eqnarray}
a\left(x\right) & = & C\exp\sqrt{2}\left[\sqrt{\frac{\alpha^{\prime}\lambda^{2}+1}{\alpha^{\prime}\lambda^{2}}}\mathbb{F}\left(x\lambda\left|-\frac{2}{\alpha^{\prime}\lambda^{2}}\right.\right)\right.\nonumber \\
 &  & -\sqrt{\frac{\alpha^{\prime}\lambda^{2}}{\alpha^{\prime}\lambda^{2}+1}}\mathbb{E}\left(x\lambda\left|-\frac{2}{\alpha^{\prime}\lambda^{2}}\right.\right)\nonumber \\
 &  & \left.-\frac{\sin\left(2\lambda x\right)}{\sqrt{\left(\alpha^{\prime}\lambda^{2}+1\right)\left(\alpha^{\prime}\lambda^{2}+1-\cos\left(2\lambda x\right)\right)}}\right],\label{eq:a solution}
\end{eqnarray}

\noindent and physical dilaton,

\begin{equation}
\phi\left(x\right)=\frac{1}{2}\Phi\left(x\right)+\frac{1}{2}\ln a\left(x\right),
\end{equation}

\noindent where $\mathbb{F}\left(\phi|m\right)$ and $\mathbb{E}\left(\phi|m\right)$
are elliptic integrals of the first and second kinds, and $C$ is
an integral constant. Now, we set $\alpha^{\prime}\lambda^{2}=16$,
$\lambda=1$, $C=1$ and plot $R\left(x\right)$, $a\left(x\right)$
and $\phi\left(x\right)$ in the region $0\leq x\leq\frac{\pi}{\lambda}$
as an example, see Fig. (\ref{fig:scalar curvature}).

\noindent In Fig. (\ref{fig:scalar curvature}), it is easy to see
that $R\left(x\right)$, $a\left(x\right)$ and $\phi\left(x\right)$
are regular in the region $0\leq x\leq\frac{\pi}{\lambda}$. To see
how $\alpha^{\prime}$ corrections affects the property of the curvature
singularity, we can see the following figure.

\begin{figure}[H]
\begin{centering}
\includegraphics[scale=0.43]{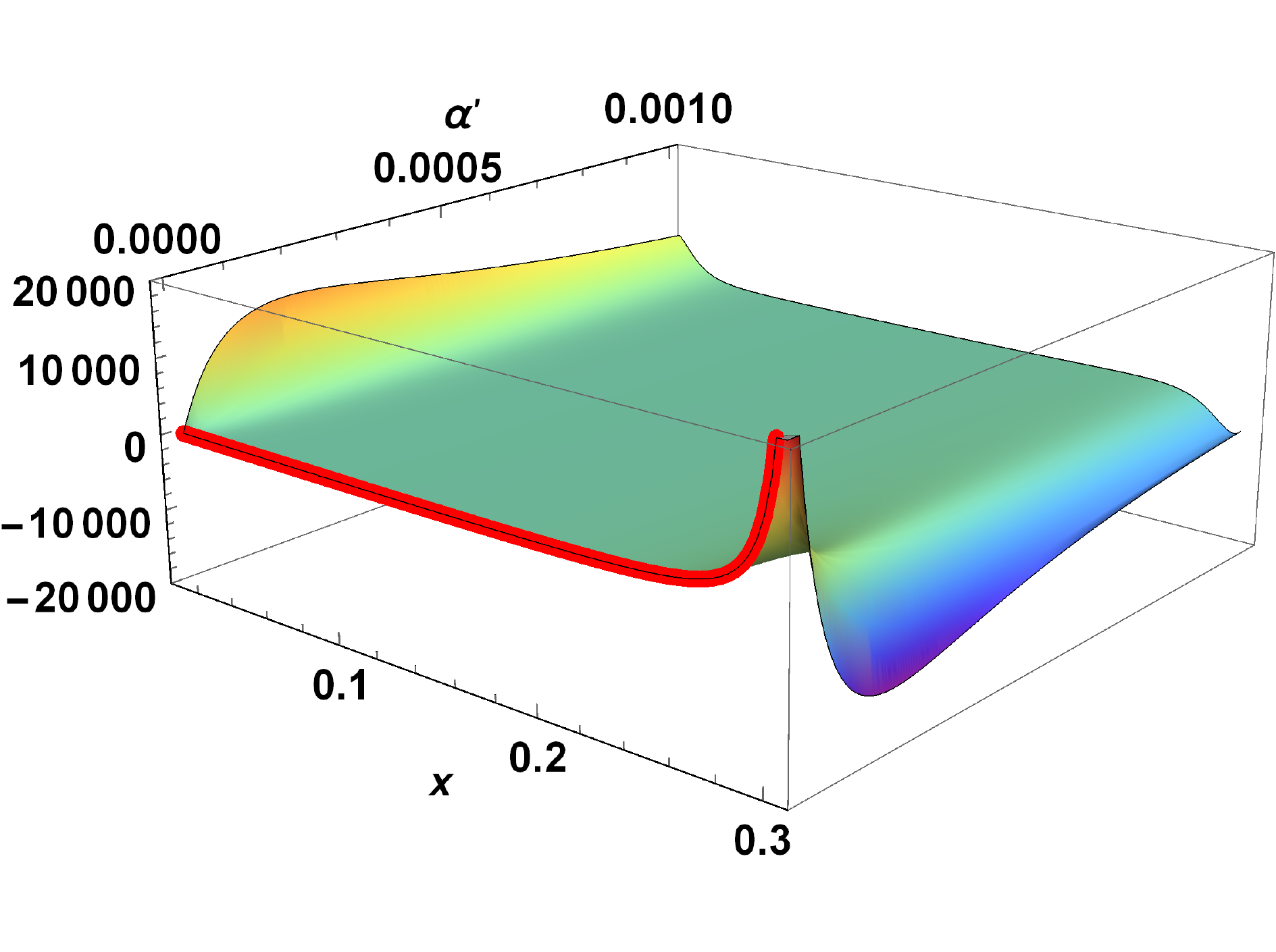}
\par\end{centering}
\centering{}\caption{\label{fig:scalar curvature all region} The figure of $R\left(x\right)$,
where $\lambda=10$, $0\protect\leq\alpha^{\prime}\protect\leq\frac{1}{1000}$
and $0\protect\leq x\protect\leq\frac{\pi}{\lambda}\simeq0.3$. The
curvature singularity locates at $x=\frac{\pi}{\lambda}\simeq0.3$
when $\alpha^{\prime}=0$.}
\end{figure}

\noindent In Fig. (\ref{fig:scalar curvature all region}), it presents
the behavior of Ricci scalar $R\left(x\right)$ of the solution (\ref{eq:solution}).
When $\alpha^{\prime}$ goes to zero, $R\left(x\right)$ reduces to
$R_{0}\left(x\right)=\lambda^{2}\cos^{-2}\left(\frac{\lambda x}{2}\right)$
(Red solid line), which possesses the curvature singularity at $x=\frac{\pi}{\lambda}$.
On the other hand, when $\alpha^{\prime}$ grows up, the curvature
singularity disappears and $R\left(x\right)$ becomes regular everywhere.

Finally, we need to stress the relations between the Dijkgraaf et
al.'s exact string black hole \cite{Dijkgraaf:1991ba} and our result.
As expected, the perturbative solutions (\ref{eq:perturbed solution})
and (\ref{eq:perturbed solution dilaton}) of Hohm-Zwiebach action
matches with $\alpha^{\prime}$ expansion of Dijkgraaf et al.'s solution
up to two-orders. The verification requires to know the higher-loop
$\beta$-function and Hohm-Zwiebach action simultaneously, and it
is not straightforward since there exits a series of field redefinitions
from the higher-loop $\beta$-function to the EOM of Hohm-Zwiebach
action. Each field redefinition modifies the EOM and their corresponding
perturbative solutions. In Appendix B, we present the simple example
of field redefinitions up to the first-order $\alpha^{\prime}$ correction.
Beyond second-order $\alpha^{\prime}$ correction, the $\beta$-functions
are unknown. It is therefore impossible to verify the correctness
of the Dijkgraaf et al.'s solution to all orders in $\alpha^{\prime}$
through the Hohm-Zwiebach action.

In short, we used the complete $\alpha^{\prime}$ corrections of closed
string theory to remove the singularity of 2D black hole or 3D black
string. We looked for the
regular black hole solution (\ref{eq:solution}) which exactly solves
the EOM of Hohm-Zwiebach action. In the perturbative limit $\alpha^{\prime}\rightarrow0$,
Witten's 2D black hole solution is recovered. The T-dual solutions can be achieved by simply replacing $a\rightarrow a^{-1}$
in eq. (\ref{eq:a solution}). Moreover, it is worthwhile to study how to obtain other exact solution (which possesses two spacetime sigularities) of coset model \cite{Sfetsos:1992yi} from the Hohm-Zwiebach action.

\begin{acknowledgments}
We thank the useful discussions with Xin Li, Peng Wang, Houwen Wu,
Haitang Yang. This work is supported  by the NSFC Grant No. 12105031.
\end{acknowledgments}

\appendix
\section{}

Supposing the coefficients $c_{k\leq n}$ are known, to figure out
the non-perturbative solutions, we can choose a different kind of
ansatz. Here, we present two simple examples. The first possible ansatz
is

\begin{equation}
\Phi\left(x\right)=-\frac{1}{2}\log\left[\sum_{\text{k=1}}^{N}\frac{\rho_{k}\left(\lambda x,c_{k}\right)}{1+\left(\alpha^{\prime}\lambda^{2}\right)^{k}}\right],\label{A:ansatz A}
\end{equation}

\noindent where $\rho_{k}$'s are functions of $\lambda x$ and coefficients
$c_{k}$. Singularities appear if and only if

\begin{equation}
\sum_{\text{k=1}}^{N}\frac{\rho_{k}\left(\lambda x,c_{k}\right)}{1+\left(\alpha^{\prime}\lambda^{2}\right)^{k}}=0.
\end{equation}

\noindent If we wish to cover the first two terms of the perturbative
solution, we set $N=2$. In the perturbative regime $\alpha^{\prime}\rightarrow0$,
the ansatz $\Phi\left(x\right)$ in (\ref{A:ansatz A}) is expanded
as

\begin{eqnarray}
\Phi\left(x\right) & = & -\frac{1}{2}\log\left(\rho_{1}+\rho_{2}\right)+\frac{\rho_{1}}{2\left(\rho_{1}+\rho_{2}\right)}\alpha^{\prime}\lambda^{2}\nonumber \\
 &  & -\frac{\left(\rho_{1}^{2}-2\rho_{2}^{2}\right)}{4\left(\rho_{1}+\rho_{2}\right){}^{2}}\left(\alpha^{\prime}\lambda^{2}\right)^{2}+\mathcal{O}\left(\alpha^{\prime3}\right).
\end{eqnarray}

\noindent To match the perturbative solution, $\rho_{k}$'s can be
fixed as

\begin{eqnarray}
\rho_{1} & = & -\frac{1}{2}\cos\left(2\lambda x\right)\csc\left(\lambda x\right),\nonumber \\
\rho_{2} & = & \frac{1}{2}\csc\left(\lambda x\right),\nonumber \\
 & \cdots
\end{eqnarray}

The second ansatz is given by

\begin{equation}
\Phi\left(x\right)=-\frac{1}{2}\log\left[\sum_{\text{k=1}}^{N}\left(\alpha^{\prime}\lambda^{2}\right)^{k-1}\omega_{k}\left(\lambda x,c_{k}\right)\right],\label{A: ansatz B}
\end{equation}

\noindent where $\omega_{k}$'s are functions of $\lambda x$ and
coefficients $c_{k}$. Singularities appear if and only if

\begin{equation}
\sum_{\text{k=1}}^{N}\left(\alpha^{\prime}\lambda^{2}\right)^{k-1}\omega_{k}\left(\lambda x,c_{k}\right)=0.
\end{equation}

\noindent In the perturbative regime $\alpha^{\prime}\rightarrow0$,
the ansatz $\Phi\left(x\right)$ in (\ref{A: ansatz B}) is expanded
as

\begin{eqnarray}
\Phi\left(x\right) & = & -\frac{1}{2}\log\left(\omega_{1}\right)-\frac{\omega_{2}}{2\omega_{1}}\alpha^{\prime}\lambda^{2}\nonumber \\
 &  & -\frac{\left(2\omega_{1}\omega_{3}-\omega_{2}^{2}\right)}{4\omega_{1}^{2}}\left(\alpha^{\prime}\lambda^{2}\right)^{2}+\mathcal{O}\left(\alpha^{\prime3}\right).\\
\nonumber
\end{eqnarray}

\noindent To match the perturbative solution, $\omega_{k}$'s can
be fixed as

\begin{equation}
\omega_{1}=\sin^{2}\left(\lambda x\right),\qquad\omega_{2}=\frac{1}{2}\cos\left(2\lambda x\right),\qquad\cdots.
\end{equation}

\section{}

Considering the FLRW ansatz

\begin{equation}
ds^{2}=-n\left(x\right)^{2}dx^{2}+a\left(x\right)^{2}dt^{2},
\end{equation}

\noindent the ordinary low energy effective action with the first-order
$\alpha^{\prime}$ correction becomes

\begin{eqnarray}
I_{o} & = & \int d^{2}x\sqrt{-g}e^{-2\phi}\left(R+4\left(\partial\phi\right)^{2}+\lambda^{2}\right.\nonumber \\
 &  & \left.+\frac{1}{4}\alpha^{\prime}R^{\mu\nu\rho\sigma}R_{\mu\nu\rho\sigma}\right)\nonumber \\
 & = & \int dxe^{-\Phi}\left[\frac{1}{n}\left(-\dot{\Phi}^{2}+H^{2}\right)+n\lambda^{2}\right.\nonumber \\
 &  & \left.+\alpha^{\prime}\frac{1}{n^{5}}\left(H\dot{n}-n\left(\dot{H}+H^{2}\right)\right)^{2}\right].
\end{eqnarray}

\noindent The corresponding perturbative solution is

\begin{eqnarray}
\Phi_{0} & = & -\log\left(\sin\left(\lambda x\right)\right),\nonumber \\
H_{0} & = & \lambda\text{csc}\left(\lambda x\right),\nonumber \\
H_{1} & = & -2\lambda^{3}\sin^{4}\left(\frac{\lambda x}{2}\right)\csc^{3}\left(\lambda x\right),
\end{eqnarray}

\noindent which is consistent with Dijkgraaf et al.'s result inside
the event horizon up to the first-order $\alpha^{\prime}$ correction.
Using the field redefinitions,

\begin{eqnarray}
n & = & n+\alpha^{\prime}\delta\mathbf{n},\nonumber \\
H & = & H+\alpha^{\prime}\delta\mathbf{H},\nonumber \\
\Phi & = & \Phi+\alpha^{\prime}\delta\mathbf{\Phi},
\end{eqnarray}

\noindent where

\begin{eqnarray}
\delta\mathbf{n} & = & \frac{1}{32}\lambda^{2}\text{\text{csc}}^{4}\left(\lambda x\right)\left(-64\cos\left(\lambda x\right)+16\cos\left(2\lambda x\right)\right.\nonumber \\
 &  & \left.+36\sin\left(\lambda x\right)-3\sin\left(3\lambda x\right)+5\sin\left(5\lambda x\right)\right),\nonumber \\
\delta\mathbf{H} & = & -\frac{1}{4}\lambda^{3}\left(4\cos\left(\lambda x\right)+1\right)\csc^{3}\left(\lambda x\right),\nonumber \\
\delta\mathbf{\Phi} & = & -\frac{1}{4}\lambda^{2}\left(\cot^{2}\left(\lambda x\right)-1\right),\nonumber \\
\end{eqnarray}

\noindent the action becomes Hohm-Zwiebach action with the first-order
$\alpha^{\prime}$ correction after setting $n=1$,

\begin{equation}
I_{HZ}=\int dxe^{-\Phi}\left(-\dot{\Phi}^{2}+H^{2}+\frac{1}{2}\alpha^{\prime}H^{4}+\lambda^{2}\right).
\end{equation}

\noindent And the corresponding perturbative solution:

\begin{eqnarray}
\Phi_{0} & = & -\log\left(\sin\left(\lambda x\right)\right),\nonumber \\
H_{0} & = & \lambda\text{csc}\left(\lambda x\right),\nonumber \\
\Phi_{1} & = & -\frac{1}{4}\lambda^{2}\left(\cot^{2}\left(\lambda x\right)-1\right),\nonumber \\
H_{1} & = & -\frac{\lambda^{3}}{4}\frac{\left(\cos\left(2\lambda x\right)+4\right)}{\sin^{3}\left(\lambda x\right)},
\end{eqnarray}

\noindent which is consistent with (\ref{eq:perturbed solution})
and (\ref{eq:perturbed solution dilaton}).
\end{document}